# Secured Data Consistency and Storage Way in Untrusted Cloud using Server Management Algorithm

Dinesh.C
Scholar, Computer Science & Engineering, MAILAM Engineering College, Mailam, Tamilnadu, India.
akdinesh85@gmail.com

## ABSTRACT
It is very challenging part to keep safely all required data that are needed in many applications for user in cloud. Storing our data in cloud may not be fully trustworthy. Since client doesn't have copy of all stored data, he has to depend on Cloud Service Provider. But dynamic data operations, Read-Solomon and verification token construction methods don't tell us about total storage capacity of server allocated space before and after the data addition in cloud. So we have to introduce a new proposed system of efficient storage measurement and space comparison algorithm with time management for measuring the total allocated storage area before and after the data insertion in cloud. So by using our proposed scheme, the value or weight of stored data before and after is measured by client with specified time in cloud storage area with accuracy. And here we also have proposed the multi-server restore point in server failure condition. If there occurs any server failure, by using this scheme the data can be recovered automatically in cloud server. Our proposed scheme efficiently checks space for the in-outsourced data to maintain integrity. Here the TPA necessarily doesn't have the delegation to audit user's data.

## 1. INTRODUCTION
Cloud computing is an internet based computing where virtual shared servers provide software, infrastructure, platform, and other resources and hosting them to customers on a pay-as-you-use basis. The reason is that whatever technologies we have known from many authors are not fully worthy and not enough to give integrity to the cloud data for its level best in cloud itself. Because they have not given full security and integrity to get satisfied to user with cloud data from CSP and it is not so efficient by using a few technologies such as Flexible distributed scheme [1] and Reed Solomon Technique [3] to save and give more integrity from modification, deletion, and append of data from cloud server. So to give yet sophisticated and efficient integrity to the cloud data so that all applications, software and highly valuable things can't be affected and modified according to someone's attempt from un-trusted cloud, we have to process and provide more security methods apart from previous methods with the help of CSP who maintains our data from cloud servers from their IP address domain [1]. But since we don't have a copy of outsourced data and data from its local site, CSP can behave unfaithfully regarding the status of our outsourced data. And since we don't have any physical data possession we have to fully depend on CSP for cloud data access without any separate login for our self access in data management. It is one of the main drawbacks that existing systems have in its module in its security level. So we have some restriction in this regard so that CSP only can take more access to manage our data with his IP address domain. But here we need to maintain data integrity for our cloud data in efficient manner for our self usage whenever there is a need for that and must have some strong security measurements for our data stored to protect from internal and external attack including byzantine as well as malicious attack. And in this case, there are some different measures such as file distribution algorithm with dynamic concept, error location correctness and data verification using distributed protocol for its integrity. But all these come from existing system with a few limitations without studying data integrity in efficient manner using cloud data weight or value. Here as proposed one, we have done storage measurement and space comparison algorithm with time management for making efficient way for data integrity. To our credit, we have done a few new implementation and they are categorized as,

(a) New implementation using the technique of Storage measurement and space comparison algorithm.

(b) Cloud storage database management with automatic time specification with the help of CSP or automatically in user's audit when user updates the cloud data.

(c) Efficient calculation of storage space before and after the data update.

(d) Server failure recovery access point in data storage and Dynamic operations in data update in each server block.

## 2. BEHAVIOR of the CLOUD
As far as cloud is concerned it has the most effective impact on internet based access for the data storage. So to maintain effective integrity and security from any internal and external attack, we must design cloud storage area in well structured manner with strong data protection against vulnerability. We can specify some unique form from cloud such as SaaS, PaaS, IaaS, and AaaS (Figure 2 explains this environment). Also Figure 1 explores the architecture of cloud and its data flow from user to the particular cloud area using TPA (but not necessarily). So to maintain this environment, we have to concentrate more on its cost effectiveness and time management. Even though there are kinds of different technologies for these atmospheres it is always a challenging task to put accurate and specified trustworthiness and cost reduction for the cloud with consistency. Cloud has the following behaviors for its credits. Moreover it can be provided for private (organization) and public usage by CSP depending on the requirements of user in their own ways without any restriction on these cloud based environments and services.

### 2.1 Keen Resources and Assurances
All resources have good assurances in multi cloud servers with the help of strong platform. The mentioned resources such as

CPU, RAM, and Bandwidth capacity in network have well formed conditions in its usage. We'll make sure that cloud server performance will never be negatively unnatural by another customer's server.

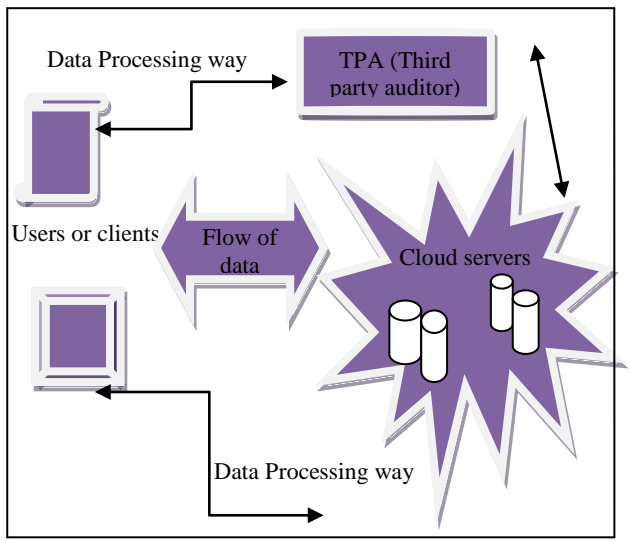

**Fig 1: Cloud Data Storage Service Architecture**

## 2.2 Surplus SAN Data Storage
All cloud server disks are mastered by a high-performance and high-availability "Storage Area Network" (SAN). Many cloud solutions operate on local disks from the host system, which means any computing or storage failure can result in down time and potential data loss. As cloud servers are independent and separate if there occur any server crack in the stored data then, these are sheltered against internal and external attack.

## 2.3 Flexibility and Durability
According to customers, they can increase and decrease the data capacity in cloud server with the help of CSP. Here storage space also has same criteria and can claim more space by on-demand request from CSP.

## 2.4 Compensate for Each Use
Customer can charge depending on the usage of network access for the data exchange with good bandwidth as well as data storage. For example, it is just like when utility company sells power to customers.

## 2.5 Effective Dynamic Computing Environment
Data needs a dynamic computing infrastructure. The basis for the dynamic infrastructure is a standardized, scalable, and secure physical infrastructure. There should be a level of redundancy to ensure high level of availability. Next, it must be virtualized.

## 2.6 Right Choice of CSP
To get excellent service from multiple servers, good service providers are important to be considered and selected. Hence utmost care must be taken in this respect so that the CSP itself can be flexible in all conditions with clients in order to get accessed with all environments (anywhere and anytime). It has the following benefits such as,

(a) Cost savings- to save the cost among IT capabilities.

(b) Reliability- data back up by CSP in cloud servers if system is stolen and loses the data itself.

(c) Scalability on demand- whenever there is a need for the data access user can be easily accessible to that anytime and anywhere in his own option.

(d) Security know-how- cloud service providers generally have more skill securing networks than individuals and IT personnel whose networks are usually connected to the public Internet anyway.

(e) Everywhere access- it allows user's ubiquitous access to computing power, storage, and related applications.

## 3. PROBLEM STATEMENT and SYSTEM FORMATION
It has the following system formation for its structure as (a) Client (b) Multiple cloud servers (c) Cloud service provider (d) Third party auditor (e) Server access point.

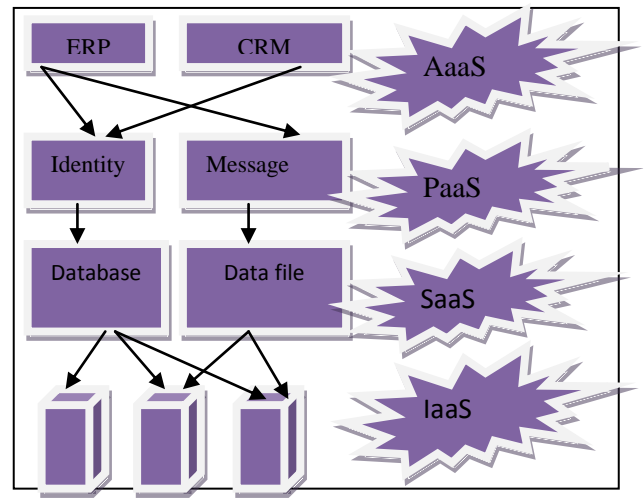

**Fig 2: Basic cloud as AaaS, SaaS, PaaS, and IaaS**

## 3.1 Putting Server Access Restore Point and Time Management
To avoid server failure in every data inclusion or any other activity by unauthorized person or any internal and external attack coming into the CSP address domain, one server access point or restore point is given to the cloud server when the client is doing some delete, modification, and append operations in his will. This is a new technique that we have introduced in our system design. The reason for which we want to do this restore point is that if there is any server failure and crash then, this restore access point helps a lot to recover everything that we have lost before that crash time itself. This process is proceeded by fixing one access point to server database (it is done with the help of CSP, because we don't have any copy of our outsourced data) when we finally finish our data exchange in cloud server. Also since we don't have any physical possession of our data in cloud server we can't have any separate login access using cryptographic key [1] for that. It is a major drawback to our cloud server to maintain our individual or group data. In the days to come, it can be rectified with the help of CSP. Here one time management is fixed in cloud database management with the help of CSP's access or automatically in server failure when user stores, deletes, modifies and appends the data and this time management is used to know the restore access point easily by CSP and clients when servers meet such failure.

### 3.2 Third Party Auditor
To save time and client's own work, TPA [2][5][9] has the authority to check the integrity of data by auditing. And TPA takes utmost care on storage correctness verification. But TPA should not have the authority to know the content of data in his independent auditing. All such audits are in the manner of privacy preserving concept so that verification can be done in privacy condition, but not public preserving condition. In the existing system, distributing protocol is used for this purpose. Here TPA has special authority for data integrity checking. However we don't give importance to TPA to such a great extent in our design but a little bit consideration only since our proposed system concentrates more on client's access.

### 3.3 Existing System
It tells us that whenever adversary is trying for any attempt to append, delete and modification for cloud data from its storage level is protected from the byzantine failure, malicious, data modification attack, and server failure using flexible distributed storage integrity auditing mechanism with distributed protocol. And it tells us that whatever changes that occur by the above mentioned failures are corrected only using specified file distribution and error correctness method. Mostly it concentrates on the TPA auditing which is always not a good option but the client for this access. And the token format is followed for data arrangement in array format for their task to be performed. All Security measurement is processed through state-of-the-art.

### 3.4 Disadvantages of Existing System
Even though it concentrates more on time consuming for user mostly the user has to depend on TPA for data integrity whenever he can't have time for auditing. Here he automatically can't have any stored data size after any such malicious, byzantine and system failures to know about data modification, delete and append in his own knowledge if there is anything. But it does not tell about its (data) fixed storage capacity for user's stored data before and after the data inclusion in cloud (how much of data has been stored in server for particular clients in his own allotted storage server area). So it is a major issue to the user and also almost users have to depend on CSP for extensive security analysis and depend partially on TPA. Also it doesn't tell effective way for server failure and doesn't make efficient way for time management concept whenever user want specified data in his earliest storage in cloud server. Here Full access is going to CSP. So CSP can behave in its own way by hiding any loss of data since existing system doesn't tell about the weight or value of stored data using any algorithm.

### 3.5 Proposed System
We assume that we know our allocated space by CSP for cloud server's access to store our data. The outsourced data has the integrity by the user by calculating weight or value of the data by our proposed method in effective manner. So in order to keep and save integrity of total storage capacity as well as client's total data storage level, we have to measure both level for our proposed goal using "storage measurement and space comparison algorithm with time management as well as server restoring access point" for the cloud collapse in some ways. In this case, we have some automatic measurement of storage space before and after the inclusion of data into the cloud using the proposed scheme when we are exchanging our data into the cloud server in need. When server failure occurs entire data may get collapsed so that user can't expect the integrity of data in its cloud level depending on some internal and external attack or CSP's action hiding the loss of data for their cost and storage purpose. Considering these drawbacks, we have designed and proposed storage measurement and space comparison algorithm as well as server restore access point to know about entire data exchange and server failure before and after the data insertion into multiple cloud server. Here the user can know if there is any modification, delete, and append operations that happen to the data from its storage level with the help of cloud database time management scheme that never has been told previously.

### 3.6 Advantages of Proposed System
By measuring storage space after and before the inclusion of data, user can know so easily the actual size of stored data before and after even though the user himself is doing any modification, deletion, and update for his own purpose by using proposed scheme. These processes are carefully done using our proposed storage space measurement and space comparison algorithm. So here user takes absolute control on the data stored in cloud apart from the TPA and we can give strong assurance to the data stored in multiple cloud server. To avoid server failure and any unexpected error, we can put one server access restore point in cloud server database for efficient data back up or restore. It is a major advantage of our proposed system. This process is done with the help of CSP for cloud database process since we don't have the physical data possession in cloud server.

## 4. EFFICIENT WAY FOR DATA STORAGE MEASUREMENT and SPACE COMPARISON ALGORITHM
Suppose that D∈S and, before data insertion, initial values are,
(a) D∉S, where D= data inserted, S is the cloud server space allocated by CSP. Now values are as the following equations,

(b) $[\sum_{i=0}^{n} [(S_i - D_i)]^{S-1}]$, where (i=>0), S= total number of multi-server and n=number of count in cloud data insertion. Now after insertion, it becomes D∈S, Here S 'server' has the data 'D' then,

(c) D={$b_i, b_{i+1}, b_{i+2}, ......, b_{i+n}$}, where b= bits of data or bytes of data or any amount of data that we add it in cloud server.

### 4.1 Data Value Measurement in Cloud Database
It is followed in a few steps when we attempt to calculate the data from its storage area excluding from CSP's allocated space. All these data insertions are checked for allocated space to test the remaining space that we want to know and to maintain integrity of data apart from some other methods such as file distribution and error correctness where existing systems have from its system design. These data calculations are clearly explained in our next part. Figure 3 explains the upload of a file to one specified web site using some software tools such as file up-loader. At present there is more such type of tools available.

## 4.2 Comparison of New Data or Data File Values with Allocated Space

Old value of the data is compared with a newly inserted data value. Hereafter, the following conditions must be satisfied with each other for the data integrity, and then in our entire data comparison, we have the following equations as,

(a) $(D \not\in S) \neq (D \in S)$ then, it is confirmed that multi-server has some data in its cloud server. Also,

(b) $[\sum_{i=0}^{n}[(S_i-D_i)]^{S-1}] + [\sum_{i=0}^{n}[(D_i)]^{S-1}]$ and the above equation is only a single time equation for overall data with its summation when client is trying for data insertion in the cloud server.

## 5. DATA BLOCK DYNAMIC OPERATIONS for OUR PROPOSED SCHEME

The following dynamic operations have to be done by user in data block update such as append, deletion, and update.

### 5.1 Append Operation

Suppose that there is 32 GB storage [5] space allocated by CSP for client's individual purpose for block length. Then initially, the allocated space is compared by storage management scheme and it is verified for its correctness or any byzantine problem by our proposed scheme by measuring the entire cloud server. Then if it is confirmed that the allocated space does not have any data in cloud server then the integrity of data is considered as strong one. Otherwise following steps have to be continued. Then the operations such as append, modification, and deletion by client are processed by space comparison algorithm for effective identification of data integrity in cloud database. So if there is any such modification by attack then, client can give assurance to the data integrity by successfully following. It is very efficient method in our design. The following algorithm explains the inclusion of data into the cloud level and here already existing storage space is taken into the account and then equations are as

(a) $[\sum_{i=0}^{n}[(S_i-D_i)]^{S-1}] + [\sum_{i=0}^{n}[(D_i)]^{S-1}]$, and

(b) $\sum_{i=0}^{n}[S_i] + [\sum_{i=0}^{n}[(T_i)]^{T-1}]$, Here, $T \not\in S$ and T are the newly added data to cloud after comparing the storage value before and after its being there. These processes are continued in a sequence manner for the client once every attempt in append operation for each block data length.

### 5.2 Deletion Operation

First, we want to compare values from existing cloud server. Then the equations defined are as

(a) $[\sum_{i=0}^{n}[(S_i-D_i)]^{S-1}] \neq [\sum_{i=0}^{n}\Phi^{[(S\Phi_i)^{\Phi i}]}]$ and Now no deletion operation is done, And on the other hand,

(b) $[\sum_{i=0}^{n}{}^{[(S\Phi_i)^{\Phi i}]} \neq \Phi]$, now the targeted data is deleted by client in successful manner. Where, $\Phi$= empty space in cloud server or no data initially.

### 5.3 Update Operation

This operation is finally finished after completion of needed action by the user. And for update, the above derived algorithms are taken into the account. For new data update, each data block should be updated automatically from already existing value to new updated value. So if we consider this data block as one array formation then, the result is as $S = (S_i \pm D_i)$ depending on the user's operation on the data. It may be positive or negative depending on the update operation to be performed.

## 6. STORAGE ACCESS POINT and TIME MANAGEMENT

To make efficient integrity of data, when user sends data to the cloud server, he can manage every time specification in its update using database management from CSP when he is doing the process. So one restore access point is made automatically in every update within the cloud database for the future purpose or to recover the data from previous stored level if there happens any data lose or server crash. When client does these operations all processes are stored in cloud database automatically with time specification accurately. To yet make efficient specification and handle this situation together without CSP's help, we will make strong protocol mechanism in the days to come.

### 6.1 Storage Algorithm Specification

1: // start and select multiple sever for data verification
2: // assume $D \in S$, where, D- data, b-data deleted and S=server,
3: If $(D \not\in S)$ // if S is not having any data
4: $S = (S_i-D_i)^{S-1}$ // for total number of server
5: else $S = (S_i+D_i)^{S-1}$
6: // Compare values, then
7: If $(D \not\in S) \neq (D \in S)$
8: $[\sum_{i=0}^{n}[(S_i-D_i)]^{S-1}] + [\sum_{i=0}^{n}[(D_i)]^{S-1}]$
9: else $\sum_{i=0}^{n}\Phi^{[(S\Phi_i)^{\Phi i}]}$ //cloud has no data
10: // Deletion operation
11: if ( $[\sum_{i=0}^{n}[(S_i-D_i)]^{S-1}] \neq [\sum_{i=0}^{n}\Phi^{[(S\Phi_i)^{\Phi i}]}]$ )
12: $S = \Phi$// no data
13: else if $[\sum_{i=0}^{n}{}^{[(S\Phi_i)^{\Phi i}]} \neq \Phi]$ // $\Phi$ means empty space and value
14: $S =$ Remove $(b_i)$ // data is deleted, b is assigned to deletion
15: return (s)
16: stop the program.

### 6.2 Storage Algorithm Detailed

(a) We should select multiple-server at a time for data auditing to know the value or weight. Then assume that $D \in S$ means that data 'D' is included in server 's' and if the condition is $(D \not\in S)$ then the server has $(S_i-D_i)^{S-1}$ space in its overall storage excluding any data addition in its space. Otherwise multiple server 'S' has addition of some data from the client in server's storage area. Then it is defined as $S = (S_i+D_i)^{S-1}$ for entire server.

(b) While doing comparison, if $(D \not\in S) \neq (D \in S)$ then, the server is calculated for any data present in its position and it is measured from the algorithm line 8 position. Otherwise it is determined that the server doesn't have any data from line 9.

(c) As far as the deletion operation is concerned if the server is not equal to the empty space then, it means that it has some data to be deleted according to user's update. It can be understood from algorithm line 11. So from that calculation, we can understand server has no data. Otherwise it can't be done. The same process again is continued by the user for entire data until when he has the access for such attempt in his own will.

### 6.3 Efficient Server Restore Access Point

Server restore access point is to be maintained from cloud storage when specified attack occurs and the size of entire cloud data is changed by modification, deletion and append operation from un-trusted cloud by such attempt so that data can lose its integrity. So this can be rectified by putting one server restore access point to the previously user's updated data from cloud data storage with time specification. So when server failure or any change occurs in the cloud storage area it is found out in TPA's auditing. And also it can be automatically corrected when the problem occurs.

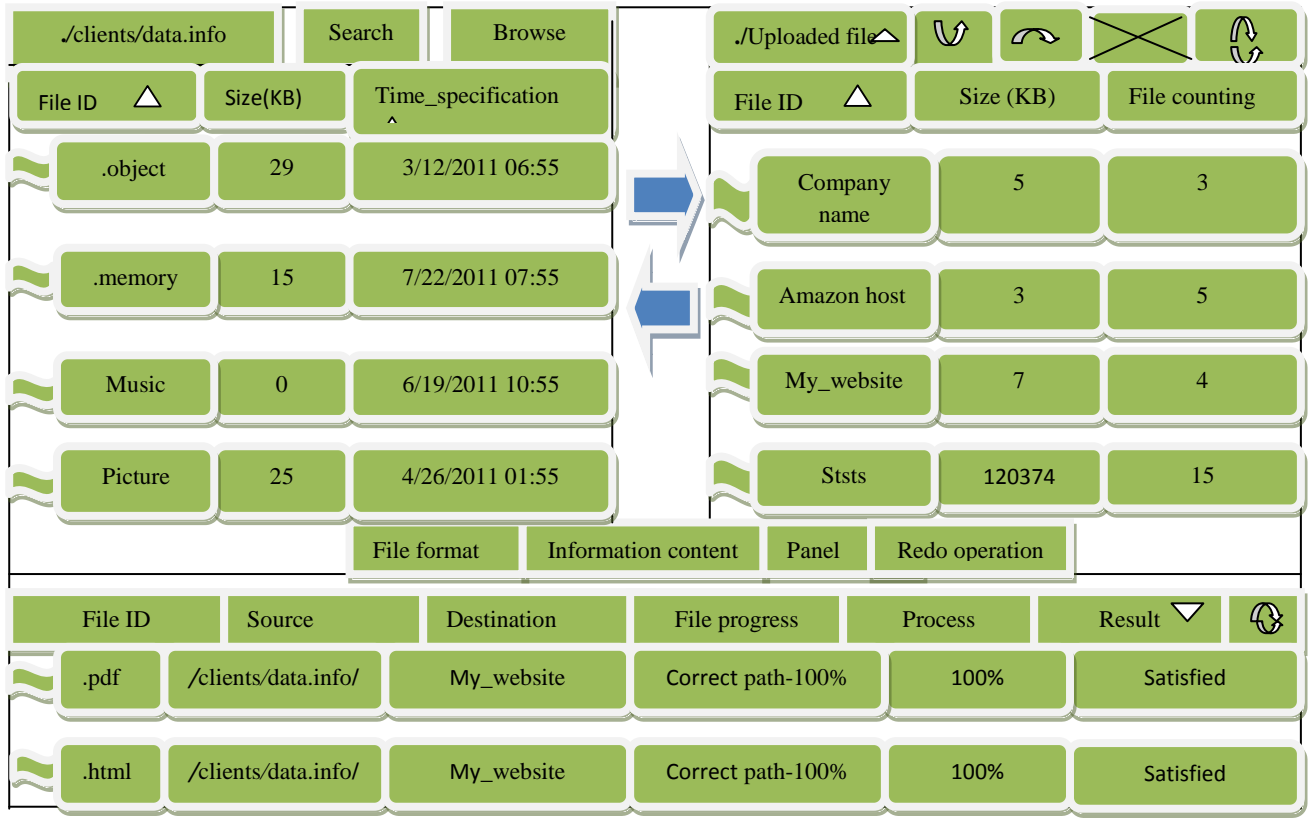

**Fig 3: Sample Cloud File Accesses and File Upload Process to Specified Site**

### 6.4 Server Restore Access Point Algorithm

1: // let x be total allocated GB for cloud data
2: // Assume that a, b and S are initial value, any data and update respectively
3: if (S ≠ X & Φ) //if server doesn't have any space
4: // No restored point is resulted in search
5: else point is found and restored
6: // then S= x+(b-a)
7: for (s=0; s<=n; s++)
8: {S=x+(b-a)}
9: S++
10: return S//continue up to each update value.

### 6.5 Restore Algorithm Detailed

For server restore access, initially select S= {$s_{i+}, s_{i+2}, s....s_{i+n}$}, for every S updated value (∀S). The empty space of cloud storage is initialized as 'a' and total allocated space is (i.e., (X-GB)) from CSP's allocated space on client's pay and request. Existing restore access point is compared with 'X' & 'Φ'. If no access point is found out then, line 3 tells that nothing is restored in its position, otherwise there exists a restore access point and it has to be restored from previous update and line 6 is processed for this. To calculate multiple server failure and for multiple restore point, line 7 is considered until all lost data are restored using previous access point from cloud data server. All these processes are followed in automatic way when server failures happen in the cloud or when client can access the cloud data or by client's request to CSP. This restore access point is most effective one yet no existing systems have specified in its concept. It can be effectively managed by clients from appropriate efficient space comparison algorithm and its restore access point algorithm.

### 7. PUBLIC VERIFIABILITY

It accesses the third party auditor [1][3][5] (as far as our paper is concerned we have not considered TPA to such a great extent for data integrity purpose since this task almost is completed by user for its full of integrity). But according to user's will, TPA may get delegated in this respect. Public are fully limited to such a great extent so that they can restrict in any operation like user.

### 8. HTTP-DoS and XML-DoS ATTACKS

This is one type of attack from the data transfer from user to cloud with the help of web-server. These attacks are corrected by back propagation of network layer and trace back algorithm. Here HTTP and SOAP protocols [4] are used for correcting these problems. So using these already existing systems, these problems can be handled effectively. These are the other types of attacks from network. This attack is trace-backed using service-oriented trace-back architecture to identify the nature of the attack from its hiding in some manner by cloud protector.

### 9. ADVANTAGES of VALUES or WEIGHT ESTIMATION from CLOUD DATABASE

While considering all other cloud concepts from different authors, here we have proposed new scheme, "storage data measurement and space comparison algorithm with time management" for data calculation to maintain the integrity of data. It is one of the efficient and different methods to protect from major data modification, deletion, and append operation in a strong way. The main purpose of our proposed scheme is that the total capacity of our allocated space after and before the date update in cloud server is accurately measured by our defined algorithm for the reason of checking how much of data we already had before any server failure or any internal and external attack and how much of data hereafter have been added into the cloud and affected by attacks after the data insertion in the cloud. We can maintain integrity of data in well secure manner using our scheme than other defined methods where the existing systems have. Figure 3 which explores cloud file accesses also explains properly the particular file upload method for its access in website. And it is purely done for client's security purpose to maintain the data in its cloud storage area and the diagram clearly tells us the time specification of uploaded file, file type, size of the file, the source where it is processed from, as well as its progress in appropriate speed. It is a specific example for file upload to the cloud storage area.

## 10. CONCLUSION

In this paper, we study the integrity and security of cloud storage data in order to give efficient proficiency to user for trustworthiness in each attempt in the data block when user is doing some updates such as append, deletion, and modification in his own will. So here user mostly has to trust CSP and TPA in some situations. In this case, CSP and TPA can behave and try for their own action so that users may lose the integrity of data. Considering this nature of atmosphere, we have proposed new technique "storage space measurement and comparison algorithm with time management as well as server recovery access point" to fix previous update for the data recovery. This is a new way that has been handled in this paper and by doing data calculation of present and previous accessed time, we can manage integrity of data from the results in user's own attempt without existing protocol and TPA support. This gives strong trustworthiness to the user over the cloud data storage where separate login can't be maintained for user's individual purpose to check integrity of data and where cryptography concept alone can't be implemented. Here byzantine, internal, external and malicious attacks are found out by proposed algorithm in efficient manner by data calculation (i.e., since there can be some changes in data) if they take any attempt on cloud storage data. The main theme of our proposed scheme is to maintain the integrity of our cloud data from specific attack [1], and also from any modification, deletion, append and malicious as well as byzantine problem since our scheme checks and tells well how much of data bits or any amount of size like gigabytes or terabytes from user's usage from CSP's allocation have been affected or modified by such attempt or problem. So user will be able to identify the nature of the cloud data update. Also it can be efficiently rectified using server restore access point from previous updated level with time management.

## 11. FUTURE ENHANCEMENT

Thus we have told our way of maintaining the data in cloud server by implementation of our proposed scheme "efficient storage measurement and space comparison algorithm with time management" as well as "server restore access point" in a well manner and we have proved right all such schemes by our appropriate well designed algorithmic specification for our proof in a clear way. There is more scope for future enhancement to get processed in maintaining security and integrity of the data using "effective read and write protocol" for the data calculation from cloud data storage in the days to come so that user can identify the attempt of different data having same weight entering into the un-trusted cloud server. This "read and write protocol" can be so efficient in identifying any data modification and other such attempt what already we have mentioned as attack in the cloud data to be stored into the server and this protocol will be working with accurate reading and protecting capacity automatically when such attempt is made by unknown person or other such attack. In our future study, we also have planned to implement and design the file locking protocol in cloud data storage point with the help of CSP for data storage and it can give clear and expected integrity level for the data in the cloud area where it is stored. And finally user can be satisfied with his requirement from cloud level with strong nature of integrity of his own data with good maintenance through the cloud service provider at multiple server as well as security of the stored data for its trustworthiness. Thus finally our next design module for the data integrity using "distributed file locking protocol as well as read and write protocol" will be hopefully so strong in maintaining integrity as well as security in cloud storage server with strong protocol protection.

## 12. REFERENCES

[1] Towards Secure and Dependable Storage Services in Cloud Computing Cong Wang, Student Member, IEEE, Qian Wang, Student Member, IEEE, Kui Ren, Member, IEEE, Ning Cao, Student Member, IEEE, and Wenjing Lou, Senior Member, IEEE-2011.

[2] A privacy preserving remote data integrity checking protocol with data dynamics and public verifiability" Qian Wang, Student Member, IEEE, Cong Wang, Student Member, IEEE, Kui Ren, Member, IEEE, Wenjing Lou, Senior Member, IEEE, and Jin Li " MAY- 2011.

[3] Enabling Public Verifiability and Data Dynamics for Storage Security in Cloud Computing Qian Wang1, Cong Wang1, Jin Li1, Kui Ren1, and Wenjing Lou2-Springer-Verlag Berlin Heidelberg- 2009.

[4] Cloud security defence to protect cloud computing against HTTP-DoS and XML-DoS attacks, Ashley,YangXiangn, WanleiZhou, AlessioBonti, Elsevier- 2010.

[5] A Multiple-Replica Remote Data Possession Checking Protocol with Public Verifiability- Second International Symposium on Data, Privacy, and E-Commerce- 2010.

[6] Addressing cloud computing security issues. Future generation computer systems-www.elsevier.com/locate/fgcs-2011.

[7] Identifying the security risks associated with governmental use of cloud computing Scott Paquette, Paul T. Jaeger, Susan C.Wilson-www.elsevier.com/locate/govinf-2010.

[8] Data Integrity Proofs in Cloud Storage-, Saravan Kumar.R, Software Engineering and Technology labs Infosys Technologies Ltd Hyderabad, India-2011.

[9] Exploiting Dynamic Resource Allocation for Efficient Parallel Data Processing in the Cloud Daniel Warneke and Odej Kao-JANUARY- 2011.